# Automated CFD shape optimization of stator blades for the PediaFlow pediatric ventricular assist device

Mansur Zhussupbekov[1], Greg W Burgreen[2], Jeongho Kim[3], and James F Antaki[1]*

[1] Meinig School of Biomedical Engineering, Cornell University, Ithaca, NY, USA
[2] Center for Advanced Vehicular Systems, Mississippi State University, Starkville, MS
[3] Department of Biomedical Engineering, Daejeon Institute of Science and Technology, Daejeon, Korea
* Correspondence: antaki@cornell.edu

**Abstract**

PediaFlow is a miniature mixed-flow ventricular assist device for neonates and toddlers. PediaFlow has a fully magnetically levitated rotor which improves biocompatibility, but the increased length of the rotor creates a long annular passage where fluid energy is lost. Therefore, a set of helical stator blades was proposed immediately after the impeller stage to remove the swirling flow and recover the dynamic head as static pressure. Automated computational fluid dynamics (CFD) shape optimization of the stator blades was performed to maximize pressure recovery at the operating point of 1.5 LPM and 16,000 RPM. Additionally, the effect on hemolysis and thrombogenicity were assessed using numerical modeling. The optimization algorithm favored fewer blades of greater length over a larger number of short blades. The ratio of wrap angle to axial length emerged as a key constraint to ensure the viability of a design. The best design had 2 blades and generated 73 mmHg of pressure recovery in an isolated stage. When re-introduced to the CFD simulation of the complete flow path, the added stator stage increased the pump head by 46% and improved the pump efficiency from 21.9% to 25.7% at the selected operating point. Automated CFD shape optimization combined with in silico evaluation of hemocompatibility can be an effective tool for exploring design choices and informing early development process.

**Keywords:** PediaFlow, pediatric, ventricular assist device (VAD), stator, optimization, computational fluid dynamics (CFD), hemolysis, thrombosis

## Introduction

Computational fluid dynamics (CFD) has been gaining popularity as a design tool for blood-wetted medical devices, including rotodynamic blood pumps such as left ventricular assist devices (VADs). This is fortuitous for determining some of the subtler features of the flow path that are not readily prescribed by traditional, textbook formulae and empirical guidelines. For example, the choices of hub and shroud profiles, blade wrap, number of blades, length of blades, volute dimensions, etc. are complex, interrelated, and affect both performance and hemocompatibility. CFD simulation can be used to explore these free parameters manually, by intuition and trial and error.[1,2] The inverse design method can facilitate this process by computing aspects of the flow path based on a specified pressure distribution.[3–5] However, for the most part, the process is labor intensive, and limited by time, resources, and the mental capacity of the designer to manage and tradeoff a multitude of parameters. Consequently, the end result is unlikely to be the truly optimal design.

This imbroglio has motivated our group to incorporate automated shape optimization for VAD design,[6–10] inspired by the success of such methods in aerospace, automotive, and other industries.[11,12] CFD-driven optimization has been recently applied to design of several VADs with encouraging results.[13–15] This approach allows exploring the design space with minimal human intervention to determine the best set of design parameters, subject to specified constraints, to maximize or minimize a quantitative objective function (or functions).

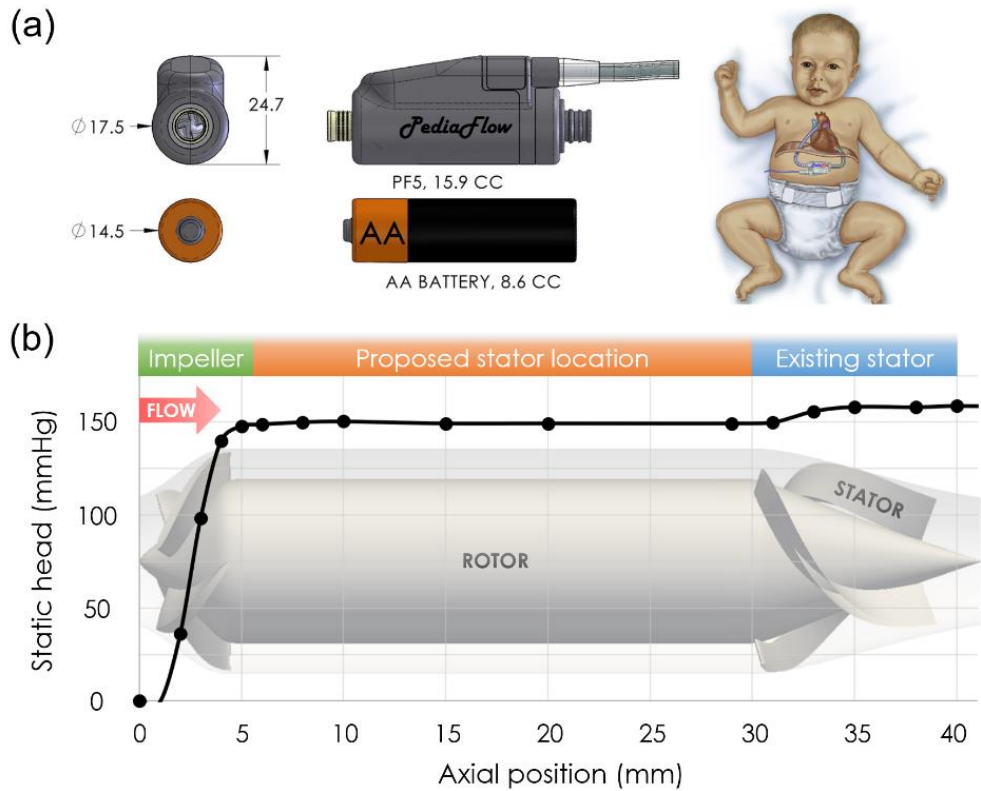

Figure 1. (a) The fifth-generation PediaFlow VAD dimensions (mm) compared to an AA-battery, designed for children from birth to 2 years of age. (b) Static head produced by PF5 at 1.5 LPM and 16,000 RPM plotted along the axial position. The image of the current PF5 rotor and stator is overlaid on the plot. The proposed mid-stator location is indicated in orange.

PediaFlow® is a pediatric, magnetically levitated, rotodynamic VAD intended to provide chronic circulatory support of children with congenital or acquired heart failure.[16,17] The fifth-generation device (PF5), shown in Figure 1 (a), is approximately the size of an AA-battery and can provide flow within the range of 0.5 − 4.0 L/min.[18] Using the inverse design method and CFD, Wu et al. (2022) increased the flow capacity and improved efficiency of the previous generation pump by increasing the inflow and outflow diameters and further optimizing the blood flow path.[19] Figure 1 (b) shows the static pressure in PF5 at 1.5 LPM and 16,000 RPM, plotted along the axial position. The flow path of the PediaFlow consists of a tapered cylindrical impeller with 4 blades on the conical inlet face, leading to a single 1.5-mm annular fluid gap region, thence passing through a 3-vane stator stage in its aft-housing.[16,20,21] Kinetic energy imparted by the impeller develops dynamic head which is then recovered (partially) as static pressure in the diffuser (stator). However, blood exiting the impeller stage must traverse a long annular path before reaching the stator, which incurs energy loss – compounded by the development of Taylor vortices within the annular path.[22,23]

Consequently, to increase pressure recovery and hence efficiency, we considered introducing a second set of stationary stator blades immediately following the impeller stage. (See Figure 1 (b).) This proposed stator stage, hereafter called *mid-stator*, is intended to remove the swirling flow and convert dynamic head into static pressure. We employed automated CFD-driven shape optimization to determine the best number of blades and optimal blade shape to maximize the pressure recovery.

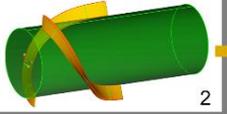

Figure 2. Input design parameters used in the optimization study shown on the isolated stator stage. Blades, in yellow, are stationary, while the rotor surface (hub), in green, spins at 16,000 RPM. The hub and shroud radii are 4 and 5.5 mm, respectively. The gap between the blade tip and the rotor is 0.1 mm.

## Materials and Methods

### *Shape parametrization*

The straight portion of the annular passage containing the helical stator stage was isolated from the rest of the flow path as shown in Figure 2. The blade geometry was fully parametrized using a B-spline in the Upfront CAE system CAESES® (Friendship Systems AG, Potsdam, Germany). The number of blades was permitted to vary from 2 to 6, and the blade shape was defined by 4 variables: axial length, wrap angle, trailing edge (TE) angle, and the area under the camber curve, controlled by a scaling factor, hereafter called the *fullness scaling factor*.

The blade leading edge (LE) was placed 2 mm from the inlet and the gap between the blade tip and the rotor was fixed at 0.1 mm. The blade LE angle, defined as the inclination of the camber line of the blade in the meridional plane with respect to the plane perpendicular to the axis of rotation, was fixed at $\alpha_1 = \tan^{-1}\left(\frac{|\bar{u}_{mer}|}{|\bar{u}_{circ}|}\right) = 4.01°$, where $\bar{u}_{mer}$ and $\bar{u}_{circ}$ are the average meridional and circumferential velocities specified at the inlet boundary.

### *Numerical solution*

The CAESES module was coupled with OpenFOAM for meshing and CFD simulation.[24] STL files created by the CAESES were imported into OpenFOAM's *snappyHexMesh* meshing utility that generated a 3-dimensional mesh comprised of hexahedra and split-hexahedra. First, a cylindrical background mesh was created using a *blockMesh* script. The STL geometry was overlaid onto the background mesh and its outlines were identified using *surfaceFeatures* utility. Next, the *snappyHexMesh* utility created a castellated mesh and refined the cells close to surfaces and sharp edges of the original geometry by cell splitting. Increased levels of mesh refinement were applied at the stator blade surface and along the leading and trailing edges and the blade tip gap. Finally, the cells were snapped to the STL surface, and additional layers of hexahedral cells aligned to the boundary surface were added in the last stage. Although using a periodic segment of the flow path is a common strategy for reducing the computational cost, creating the cyclic boundary patches and subsequent mesh generation often failed in the extreme cases of blade design. Therefore, to ensure robust execution of the parametric CAD generation and meshing scripts, the full flow path was simulated. The mesh size varied from 2.5 to 3.5 million cells depending on the number of blades and the blade surface area.

The inlet velocity boundary condition (BC) was prescribed using OpenFOAM's *cylindricalInletVelocity* BC type where axial, radial, and tangential velocity components were specified as follows: a uniform axial velocity corresponding to the flow rate of 1.5 lpm, 0 radial component, and a tangential velocity corresponding to $\omega r$, where $\omega$ = 16,000 rpm and $r$ = [0.004, 0.0055] m spanning the distance from hub to shroud surfaces. The rotating wall velocity was prescribed at the hub surface and no-slip BC applied at the shroud. The pressure at the outlet of the domain was set to zero. Blood was treated as an incompressible Newtonian fluid with the density of 1060 kg m$^{-3}$ and viscosity of 3.5 cP since the shear rate levels inside the computational domain exceeded 100 s$^{-1}$. No turbulence modeling was included since the choice of a turbulence model and its parameters is often associated with user discretion and uncertainty.[25] After a steady-state flow solution was obtained, the pressure gain ΔP was calculated as the difference between the outlet pressure (zero) and the average pressure on the inlet patch.

Next, hemolysis was computed using the Giersiepen-Wurzinger power law model[26,27] employing the Eulerian approach proposed by Garon & Farinas.[28] Specifically, the asymptotically consistent formulation described in Farinas et al.[29] was implemented in OpenFOAM, using constants published by Heuser & Opitz[30,31] since they are applicable in a wider range of shear stress values and demonstrated excellent correlation coefficient in the study by Taskin et al.[32] The scalar shear stress in the hemolysis source term was computed following Faghih & Sharp.[33] Since most numerical models of hemolysis do not predict an accurate absolute value,[34] the relative hemolysis was evaluated using the Relative Hemolysis Index (RHI), similar to Gil et al.,[35] calculated as RHI = HI/HI$_{nominal}$, where HI$_{nominal}$ is the hemolysis generated by the annular section of the flow path with no stator blades at the same operating condition.

The OpenFOAM module, comprising mesh generation, flow solution, and hemolysis evaluation, was automated using a Bash script to output ΔP and RHI values, that were fed back to CAESES.

*Optimization Procedure*

The blade number and blade geometry were optimized to maximize the pressure gain ΔP (the objective function) using a two-stage algorithm illustrated in Figure 3: an exploration stage using the Sobol method[36] in which design sensitivities were evaluated and best candidates identified, followed by a local optimization of those variants using a gradient-free T-search algorithm.[37] (Formally, the objective function used for optimization was to minimize -ΔP; however, we present the study in terms of maximizing ΔP.)

Two Sobol exploration studies were performed. As a general rule, number of repeated analyses is proportional to the third power of the number of design variables.[38] In the first round of exploration, the overall dependencies and trends were evaluated using 250 design variants. The first round only included the objective function calculation (ΔP) without the hemolysis evaluation. In the second round, the input parameters and constraints were adjusted based on initial observations, and 125 design variants were assessed for pressure recovery and hemolysis. The subsequent local optimization performed on the best candidates involved 40 design iterations each, where the number of blades was kept the same and the remaining four parameters were optimized using the T-search algorithm.

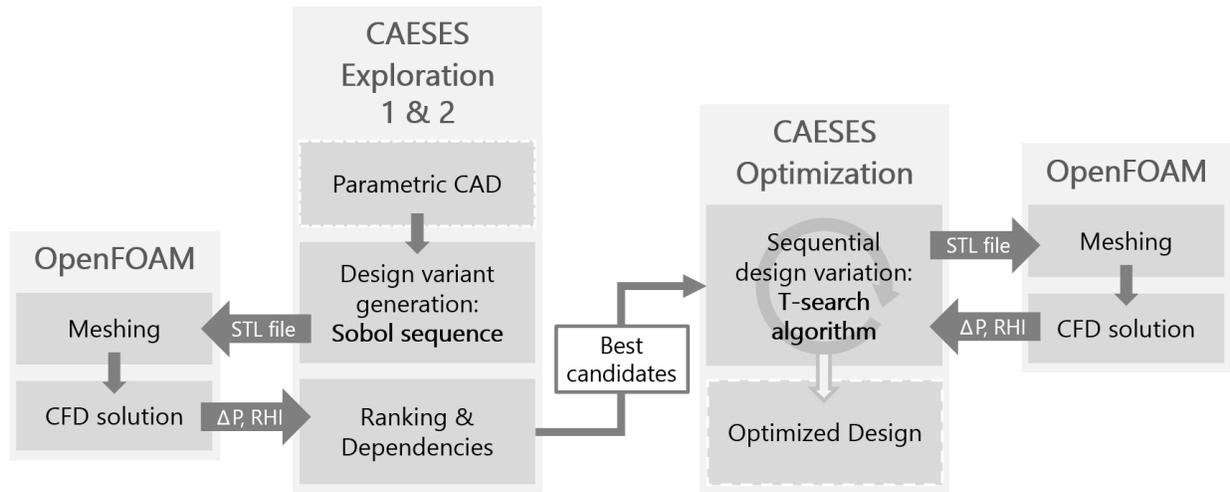

Figure 3. Schematic depiction of the optimization process and communication between the CAESES and OpenFOAM modules.

Approximate time for single design evaluation including mesh generation and CFD solution was 40 minutes. Thus, the Exploration 1 study with 250 design evaluations was completed in 7 days. The processor specifications were Intel Xeon Gold 2.4GHz, 3.7GHz Turbo, 20 cores, 2 threads per core.

*Thrombosis evaluation*

After selecting best design variants based on pressure recovery and hemolysis, a multi-constituent numerical model of thrombosis of Wu et al. was used to assess the thrombogenicity of the optimized designs.[39] Briefly, blood was treated as a multi-constituent mixture comprised of a fluid component, assumed to behave as a linear viscous (Newtonian) fluid, and a thrombus component, treated as a porous medium. The transport and reactions of 10 chemical and biological species were incorporated using a system of coupled convection-reaction-diffusion equations to represent platelet activity and thrombus formation. Biomaterial-related parameters in the model were previously calibrated for the Titanium alloy Ti6Al4V, the blood-contacting material of the PediaFlow.[40] Applying the same numerical setup, a thrombosis simulation was conducted assuming whole blood of 3.5 cP viscosity and 1060 kg/m$^3$ density. The total platelet concentration at the inlet was prescribed as $3\times10^{14}$ PLT/m$^3$ with 1% background level of platelet activation.

*Verification in the full pump*

To verify the hydraulic performance of the optimized mid-stator in the full pump, it was re-introduced to the CFD simulation of the complete flow path; the resulting pressure head and efficiency were then compared to the original pump. The distance from the impeller blade TE to the mid-stator blade LE was 1.3 mm. Meshing was performed using *snappyHexMesh* and CFD solution was performed in OpenFOAM using the frozen-rotor approach.

**Results**

Table 1 shows the upper and lower constraints used for the input parameters in Exploration 1 and Exploration 2. The correlation charts from Exploration 1 showing the dependence of the objective function on the input variables are presented in Figure 4 (a)-(e). In this first stage, the wrap angle had a strong influence on ΔP, however, the rest of the input variables demonstrated a weak dependence. Designs that produced pressure loss instead of pressure gain (not included in the charts) were characterized by extreme values of the wrap angle to axial length ratio. Excessively low ratios produced blades did not provide sufficient flow guidance, resulting in flow separation, whereas excessively high ratios produced blades that were too congested, resulting in flow blockage. When plotting the objective function against the wrap angle to axial length ratio, shown in Figure 4 (f), the resulting correlation had a notable jump in the range between 10 and 30 degree/mm.

Table 1. Design input variables and their constraints used in Exploration 1 and 2.

| Input parameter | Exploration 1 | | Exploration 2 | |
| --- | --- | --- | --- | --- |
| | Lower constraint | Upper constraint | Lower constraint | Upper constraint |
| N blades | 2 | 6 | 2 | 4 |
| Axial length | 6 mm | 19 mm | 5 mm | 19 mm |
| Wrap angle | 30° | 240° | — | — |
| TE angle | 70° | 90° | 50° | 90° |
| Fullness scaling factor | 0.95 | 1.2 | 0.95 | 1.2 |
| Wrap angle to axial length ratio [degree/mm] | — | — | 10 | 30 |

For the reason above, the blade wrap angle in Exploration 2 was prescribed using the wrap angle to axial length ratio and the values were constrained to the favorable range. The constraints for other variables were adjusted as well. Notably, since designs with fewer blades performed better in Exploration 1, the upper constraint for the number of blades was reduced from 6 to 4. The number of blades was previously sampled by the Sobol design engine as an integer variable, which resulted in fewer designs at the lower and upper ends of the range, as can be seen in Figure 4 (a). To circumvent this issue in the second exploration stage, the number of blades was prescribed as a continuous variable where the actual number was the integer portion of the value, which resulted in a more uniform sampling of the design space, shown in Figure 5 (a). Additionally, hemolysis evaluation was included in this second exploration stage.

The results of Exploration 2 are shown in Figure 5. Within this smaller subset of the fuller design space, trends from the previous stage continued except for the axial length, which changed from a slightly negative to a strong positive correlation. Hemolysis evaluation revealed that RHI increased with greater number of blades as well as axial length and wrap angle. (See Figure S1.) Figure 6 shows the plot of relative hemolysis index, RHI, against the pressure recovery, ΔP. A non-inferior envelope can be traced along the lower edge of the point cloud, shown with a dashed line, denoting the lowest hemolysis obtained for given pressure recovery. From this plot, two design variants with 2 and 3 blades, producing 70 mmHg and 63 mmHg of pressure recovery, respectively, were selected for further local optimization.

As the final stage of optimization, the two best candidates were optimized using a local T-search algorithm. The evolution of the objective function from the initial to optimized design points is given in Figures S2 and S3; the final values of the input variables for the optimized designs are given in Table S1. The velocity fields, RHI, and thrombosis are shown in Figure 7. The velocity vector plots indicate well-attached flow for both the 2- and 3-blade designs. As a result of local optimization, the best 2-blade candidate increased ΔP by 3.7% to 72.7 mmHg but at the cost of 2.5% increase in RHI to 2.06. The best 3-blade candidate improved ΔP by 1% to 63.6 mmHg with negligible increase in RHI by 0.4% to 1.86. After 30 minutes of thrombosis simulation, the 2-blade design developed 3D thrombi on the suction side of the blades at the blade-shroud interface towards the trailing edge. In contrast, the 3-blade design remained free of thrombus at same time point.

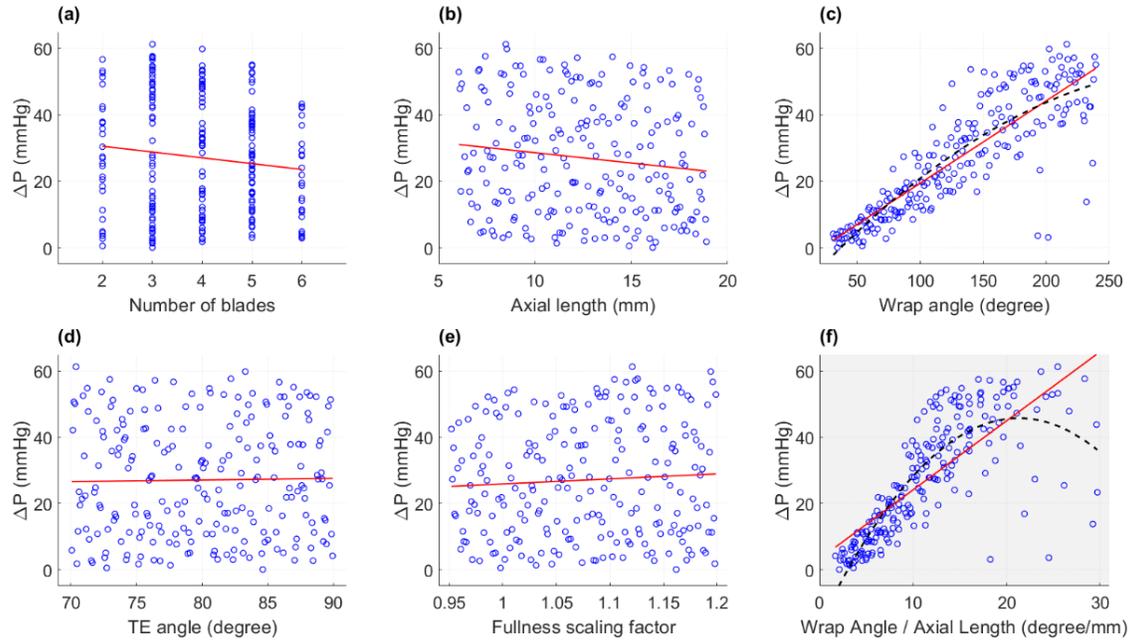

Figure 4. Results of Exploration 1. (a)-(e) Pressure recovery, ΔP, plotted against input variables with 1st-order fit plotted in solid red and 2nd-order fit shown in dashed line. Only designs that produced pressure rise are shown. (f) Pressure recovery plotted against the ratio of wrap angle to axial length showed a sharp increase in the range between 10 and 30 degree/mm.

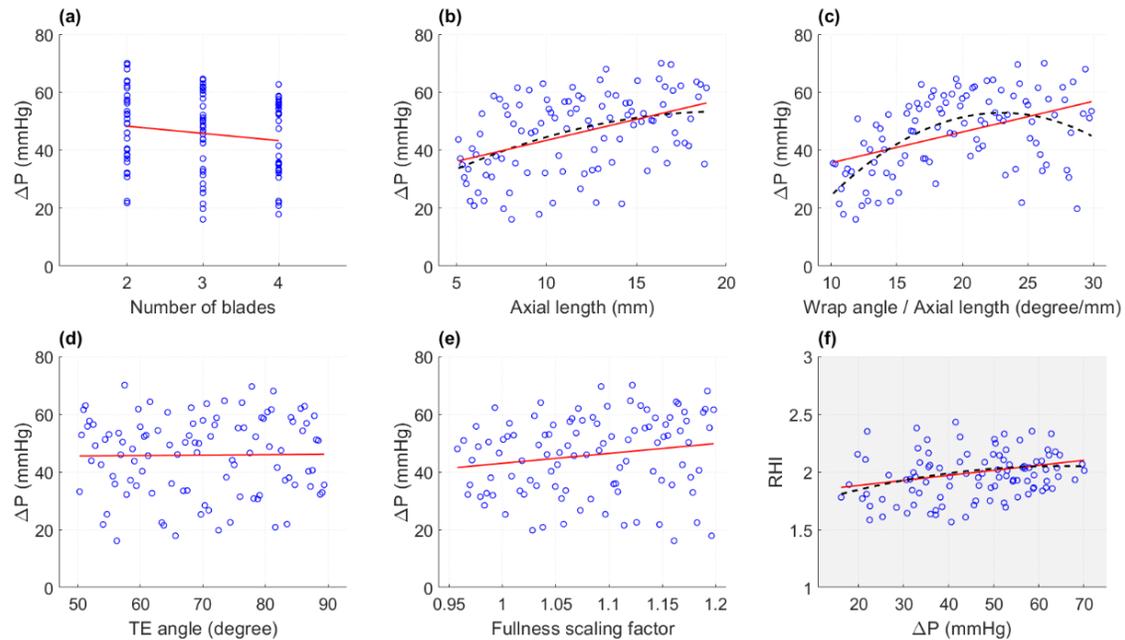

Figure 5. In Exploration 2, the wrap angle input parameter was replaced with the ratio of wrap angle to axial length, constrained to the favorable range identified from Exploration 1. (a)-(e) Pressure recovery, ΔP, plotted against input variables with 1st-order fit plotted in solid red and 2nd-order fit shown in dashed line. (f) The relative hemolysis index, RHI, plotted against ΔP reveals that pressure recovery came at the cost of increased hemolysis.

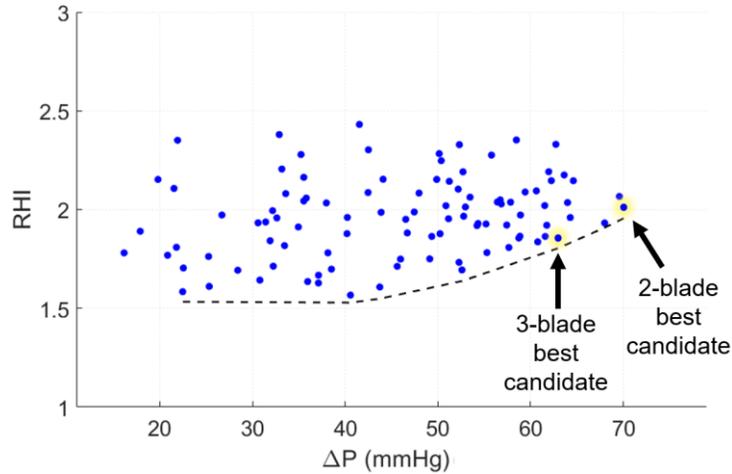

Figure 6. The relative hemolysis index, RHI, plotted against ΔP from Exploration 2. Dashed line traces the outer envelope of designs with the lowest RHI for given ΔP. Arrows point to the two best candidates selected for further local optimization.

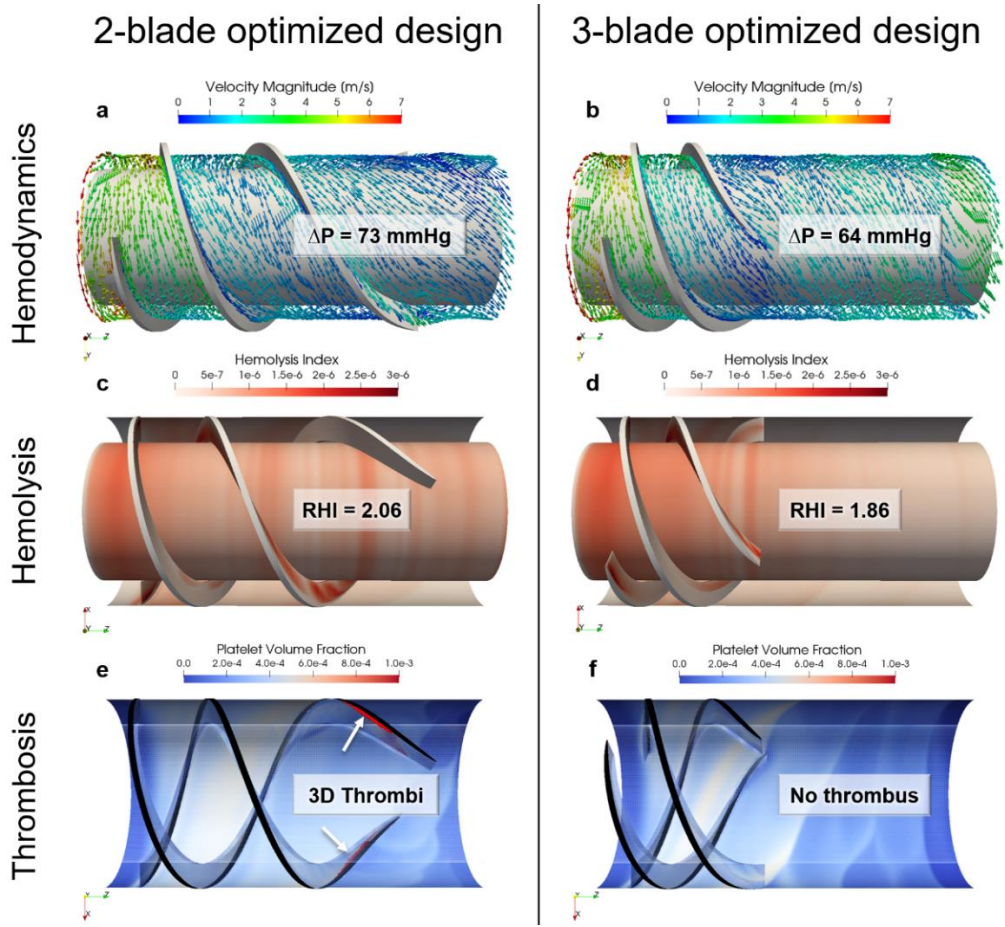

Figure 7. Comparison of optimized 2-blade and 3-blade designs. Flow is from left to right. (a, b) Velocity vectors presented at mid span. (c, d) RHI shown on blood contacting surfaces. (e, f) Thrombosis simulation results shown at 30 min of simulation time. The deposited platelet volume fraction is shown on blade and shroud surfaces; threshold volume fraction for 3D thrombus visualization is > 0.1. The hub surface remained free of thrombus in both cases.

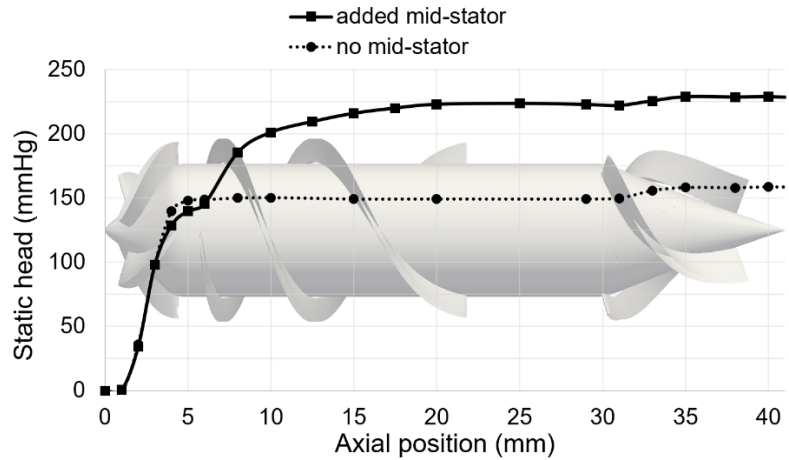

Figure 8. Performance of the optimized 2-blade design in the PF5 pump at 1.5 LPM and 16,000 RPM compared to the baseline. The image of PF5 with added mid-stator is overlaid on the plot to match the axial position; original PF5 not shown.

The full PF5 pump fitted with the 2-blade optimized mid-stator produced 218 mmHg of pressure head at 1.5 LPM, 16,000 RPM, constituting a 69 mmHg increase over the original PF5. Figure 8 shows the static head plotted along the axial position for the modified and original pumps. As can be seen, the added mid-stator stage adversely affects the pressure within the trailing region of the impeller stage. Subsequently, the head increases by approximately 80 mmHg within the mid-stator stage, after which it remains relatively flat, followed by an additional rise across the aft stator stage. The added mid-stator stage improved the pump efficiency from 21.9% to 25.7%.

**Discussion**

VADs are unique from traditional turbomachinery by virtue of their requirement for miniaturization and biocompatibility. PediaFlow's iterative design evolution over the years yielded a miniature design comparable to AA-battery in size. The fully magnetically levitated rotor eliminated mechanical bearings, a potential nidus for thrombus formation. This improves biocompatibility but requires an increased rotor length, which results in an unorthodox design with a long annular space between impeller and stator. The problem of designing an additional stator stage in this scenario lends itself well to automated optimization since it can help gain insight when little experience is available or when previous experience is not directly applicable. Design exploration using optimization can also provide options from which to choose, complementing the knowledge base for decision making.[37]

The exploration studies reported here revealed the most influential input parameters determining ΔP were the axial length, wrap angle, and the number of blades. (See Figure 4 and Figure 5.) Crucially, the ratio of wrap angle to axial length emerged as a key constraint to ensure the viability of a design. In Exploration 1, all design variants that produced more than 40 mmHg of pressure rise had the wrap angle to axial length ratio between 10 and 30 degree/mm (Figure 4 (f)). The change of sensitivity slope of Figure 4 (b) to Figure 5 (b) demonstrates the non-linearity of the five-dimensional design space. When a slightly smaller subset (Exploration 2) is considered of the fuller (Exploration 1) design space, the more local and inversely different relationship of pressure generation to axial length is revealed.

Surprisingly, ΔP demonstrated a weak dependence on the TE angle, as can be seen from Figure 4 (d) and Figure 5(d). This could be attributed to the fact that the rotor imparts considerable circumferential velocity on the fluid through viscous forces before it exits the annular gap section, so the benefit of completely removing the swirling flow would be lost.

Similar to the hydraulic performance, hemolysis was most affected by the number of blades, blade axial length, and wrap angle. (See Figure S1.) Greater number of blades and longer blades produced more red blood cell damage. Interestingly, while the RHI demonstrated low sensitivity to the ratio of wrap angle over axial length, reconstructing the wrap angle values in Figure S1 (f) shows a strong positive correlation. From Figure 6, hemolysis was a competing function with pressure recovery. While minimizing hemolysis was not included as an objective function, the lower outline describing the pressure-hemolysis tradeoff is reminiscent of a Pareto front (non-inferior set) encountered in multi-objective optimization and can help avoid sub-optimal designs that produce excessive hemolysis for given pressure rise.

Consistent with the trends above, local optimization of best candidates yielded slight improvement in pressure recovery albeit at the cost of a concomitant increase in hemolysis. Application of the numerical model of thrombosis to the two optimized designs offered another point of comparison and revealed potential areas of concern. The longer design variant with 2 blades developed 3D thrombi on the suction side of blades at the blade-shroud interface. Adding a fillet to the base of the stator vanes could be explored to improve washing and thromboresistance (in addition to manufacturability).

It is important to acknowledge limitations of this study. Most notably, the study focused on a single operating point, therefore off-design operation conditions must be considered if the proposed mid-stator is to be implemented in a clinical device. A multi-point optimization with weighted-sum approach could be applied in which the objective function encompasses the pump performance across a wider operating window. Also, these simulations assumed steady flow, whereas in the body the flow will be pulsatile due to the contribution of the native ventricle. For economy of computational cost, the mid-stator stage was considered in isolation, hence used an idealized inlet BC. The consequence of neglecting the influence of the stator on the trailing region of the impeller was to overestimate the overall pressure output. In subsequent work, optimization must be performed in conjunction with the impeller and existing stator stages and could be combined with the inverse design approach to inform the parametric design and relevant constraints. Critically, the improved hydraulic performance must be carefully weighed against the added risk of thrombogenicity. Conventional wisdom dictates that it is best to have an unobstructed flow with good washing of blood-contacting surfaces. It must be mentioned that the numerical model of thrombosis used in this study has been calibrated using in vitro microfluidic experiments but not against full-scale VAD experiments. Finally, manufacturability was not considered as a constraint in this study.

In summary, this study applied automated CFD shape optimization to a pediatric VAD and demonstrated the utility of the method for exploring and evaluating ideas, generating design choices, and informing development process. The resulting mid-stator stage significantly improved overall pressure recovery and efficiency of the PediaFlow PF5 VAD. In silico evaluation of hemolysis and thrombogenicity revealed competing objectives and exposed possible hemocompatibility issues, contributing to early decision making.

**Acknowledgments**

This work was supported by the National Institutes of Health grant R01HL089456 and the U.S. Army Medical Research Acquisition Activity Project Number W81XWH2010387. We would also like to thank Friendship Systems AG for providing a free academic license for CAESES. Lastly, we are grateful to Dr. Jingchun Wu (Advanced Design Optimization, LLC, Irvine, CA) for sharing his invaluable experience in VAD design and optimization.

# SUPPLEMENTARY MATERIAL

Table S1. Input parameter values of the two optimized designs.

| Input parameter | 2-blade optimized design | 3-blade optimized design |
| --- | --- | --- |
| N blades | 2 | 3 |
| Axial length [mm] | 18.0565 | 9.8125 |
| Wrap angle to axial length ratio [degree/mm] | 23.8750 | 24.3223 |
| TE angle [degree] | 57.50 | 51.25 |
| Fullness scaling factor | 1.14546 | 1.11719 |

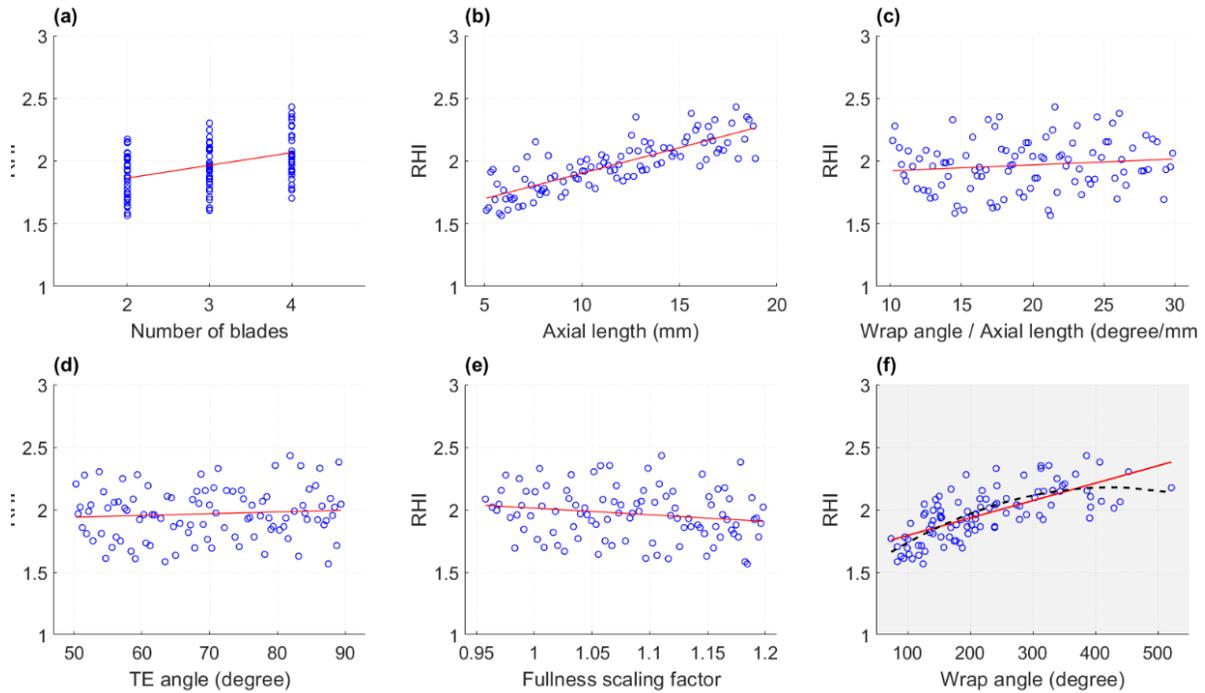

Figure S1. Sensitivity of the relative hemolysis index, RHI, in Exploration 2. (a)-(e) RHI plotted against input variables with 1st-order fit plotted in solid red. (f) RHI plotted against wrap angle (reconstructed by multiplying axial length and wrap angle / axial length ratio) with 2nd-order fit shown in dashed line.

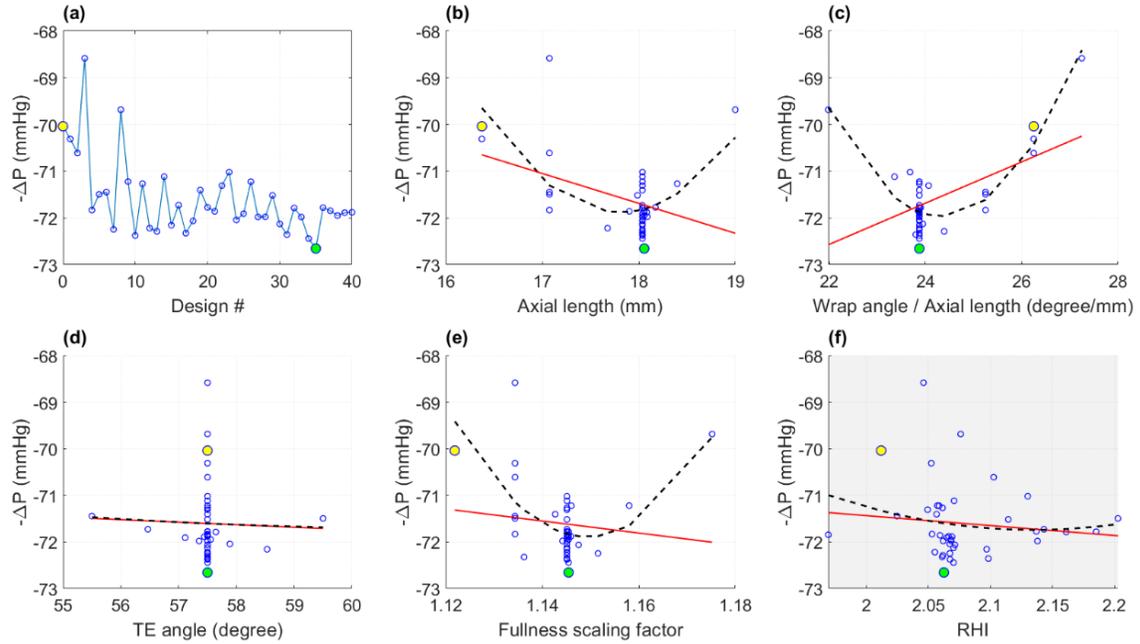

Figure S2. Local optimization of the best 2-blade candidate using the T-search algorithm. -ΔP is plotted since formally the objective function used for optimization was to minimize -ΔP. The initial design point is highlighted in yellow, and the final (optimized) design is highlighted in green. (a) The evolution of the objective function with consecutive design variations. (b)-(e) The objective function plotted against the input variables. (f) The gain in pressure recovery came at the cost of increased hemolysis, RHI.

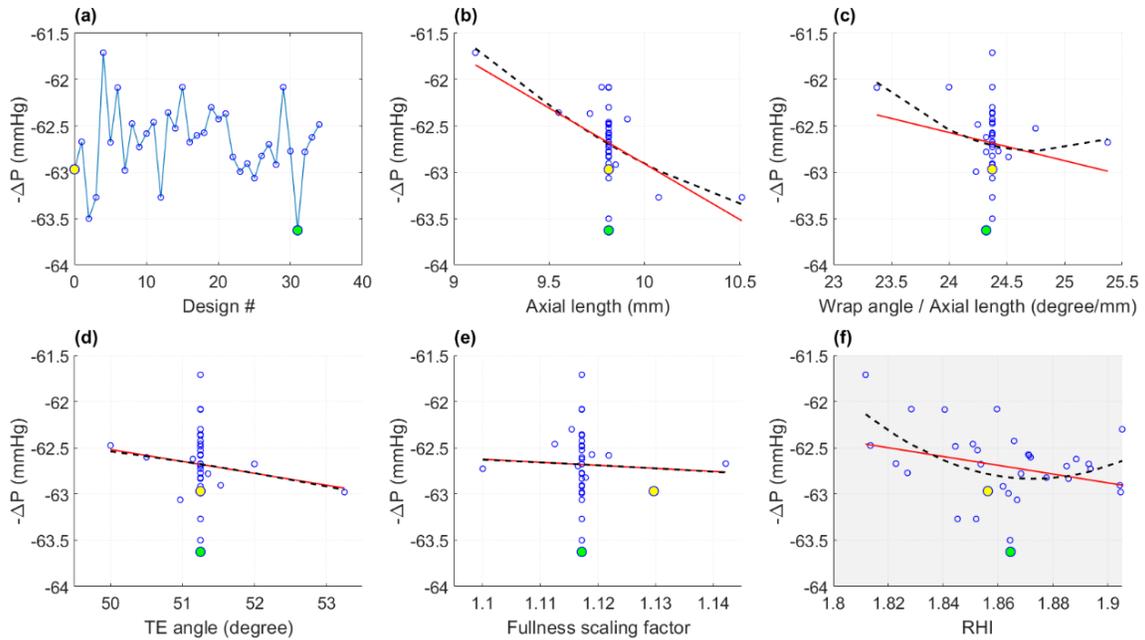

Figure S3. Local optimization of the best 3-blade candidate using the T-search algorithm. The initial design point is highlighted in yellow, and the final (optimized) design is highlighted in green. (a) The evolution of the objective function with consecutive design variations. (b)-(e) The objective function plotted against the input variables. (f) The gain in pressure recovery came at the cost of increased hemolysis, RHI.